\documentclass[11pt,twoside]{article}

\usepackage{asp2006}
\usepackage{graphics}
\usepackage{epsf}

\begin{document}
\title{On the Latitude Distribution of the Polar Magnetic Flux as Observed
by SOLIS-VSM}   
\author{N.-E. Raouafi, J. W. Harvey and C. J. Henney}   
\affil{National Solar Observatory, Tucson, Arizona, USA}    

\begin{abstract}

Magnetograms from the Vector SpectroMagnetograph (VSM) of the Synoptic Optical Long-term
Investigations of the Sun (SOLIS) project are utilized to study the latitude distribution of
magnetic flux elements as a function of latitude in the polar solar caps. We find that the density
distribution of the magnetic flux normalized by the surface of the polar cap and averaged over
months decreases close to the solar poles. This trend is more pronounced when considering only flux
elements with relatively large size. The flux density of the latter is relatively flat from the
edge of the polar cap up to latitudes of 70$^\circ$--75$^\circ$ and decreases significantly to the
solar pole. The density of smaller flux features is more uniformly distributed although the
decrease is still present but less pronounced. This result is important in studying meridional
flows that bring the magnetic flux from lower to higher solar latitudes resulting in the solar
cycle reversal. The results are also of importance in studying polar structures contributing to the
fast solar wind, such as polar plumes.
\end{abstract}


\vspace{-0.5cm}
\section{Introduction}

Polar regions of the Sun harbor numerous challenging solar phenomena, such as the source and
acceleration process of the fast solar wind. Although it is widely believed that the magnetic field
is main driver of most physical processes within the solar polar areas, these areas are not
adequately characterized for observational and instrumental limitation reasons. 

Several solar phenomena (i.e., solar differential rotation, supergranular diffusion and meridional
flows) couple together to transport poleward the mid-latitude magnetic flux of decaying active
regions. This process leads to the formation of polar caps and their evolution through the solar
cycle (see Babcock \& Babcock 1955; Leighton 1964; DeVore \& Sheeley 1987; Sheeley, Nash, \& Wang
1987; etc.). Solar meridional flows have been shown, both observationally and theoretically, to
be the main mechanism of zonal flux transport (Howard 1974; Durrant, Turner, \& Wilson 2004; Duvall
1979; LaBonte \& Howard 1982). However, the weak flow, of few times 10~m~s$^{-1}$ at best, is not
easily measurable and direct measurements can not be achieved beyond solar mid-latitudes
($\sim45^\circ$). Thus, it is important to obtain additional constraints on these flows close to
the solar poles, which would be of great use for solar dynamo and flux transport models.

Raouafi, Harvey \& Solanki (2006a,b; 2007) studied polar plumes EUV spectral emissions using
different models. By comparing theoretical results to observations, they found that polar plumes
would preferentially be based more than $10^\circ$ away from the solar pole. Saito (1958) noticed
that white-light plumes in eclipse observations were also rooted close to polar hole edges. The
study of the polar flux latitude-distribution would confirm Raouafi et al.'s findings and constrain
the magnetic flux transport (e.g., meridional flows) that drive such a distribution.

\section{Observations and Data Analysis}

\begin{figure}[!h]
 \plotone{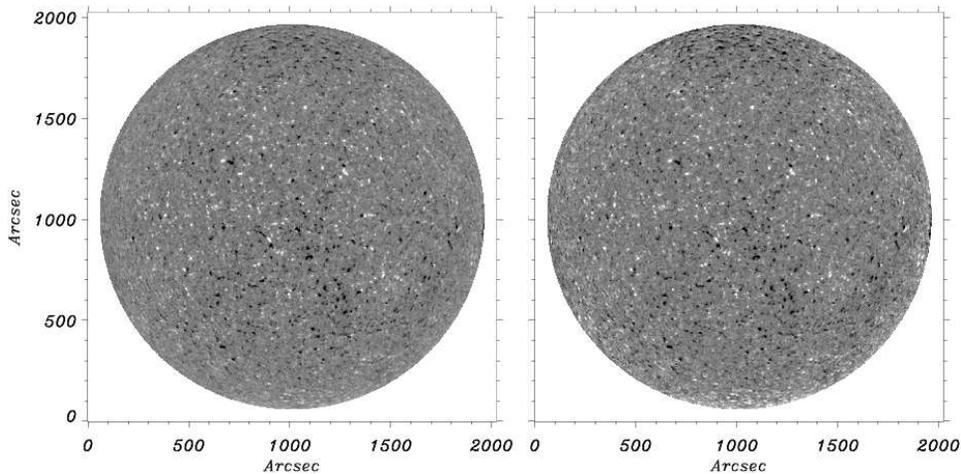}
 \caption{SOLIS/VSM-chromospheric (Ca~{\sc{ii}}~854.2~nm; Sep. 11, 2007) LOS-magnetogram with
 (right) and without (left) radial correction.
     \label{solis_8542_110907_fdisk_spw5_paper}}
 \end{figure}

Line-of-sight (LOS) chromospheric (e.g., Ca~{\sc{ii}}~854.2~nm) magnetograms from the VSM
SpectroMagnetograph (VSM; Jones et al. 2002) are utilized to characterize the latitude distribution
of magnetic flux elements in the northern polar cap during June-November 2007. The VSM is part of
the Synoptic Optical Long-term Investigations of the Sun (SOLIS; Keller, Harvey \& Giampapa 2003).
The LOS-chromospheric component of the magnetic field benefits from the canopy structure at this
height providing a strong signal near the limb (see left panel of 
Figure~\ref{solis_8542_110907_fdisk_spw5_paper}). To further improve the signal, in particular
close to the limb, a normalizing radial correction described by Raouafi, Harvey \& Henney (2007) is
applied to every magnetogram (see right panel of Figure~\ref{solis_8542_110907_fdisk_spw5_paper}).
A threshold on the field strength is subtracted from the magnetograms to suppress the noise.

Magnetic flux elements are selected by applying a top-hat operator that selects the intensity peaks
in the magnetic field strength of each polarity map. This operator uses a disk structuring element
with sizable radius to select the prominent peaks with base sizes larger than the structuring disk
element. A shape constraint can also be included in the structuring element. Weak-diffuse magnetic
fields are ignored in the selection process. Once magnetic elements of interest are identified,
their locations in terms of latitude and longitude are determined by averaging the coordinates of
the contour of each of them. The distribution of the selected elements as a function of latitude is
obtained by determining the average location for every selected feature. However, the obtained
distribution is absolute and might be biased by the latitude area dependence. In order to avoid
that, the obtained distribution are normalized by the latitude area distribution taking into
account the solar geometry (i.e., $B_0$ angle). Since single histograms do not show clearly the
distribution of magnetic flux elements due to statistical reasons, they are monthly averaged. 

\section{Results}

 \begin{figure}[!h]
 \plotone{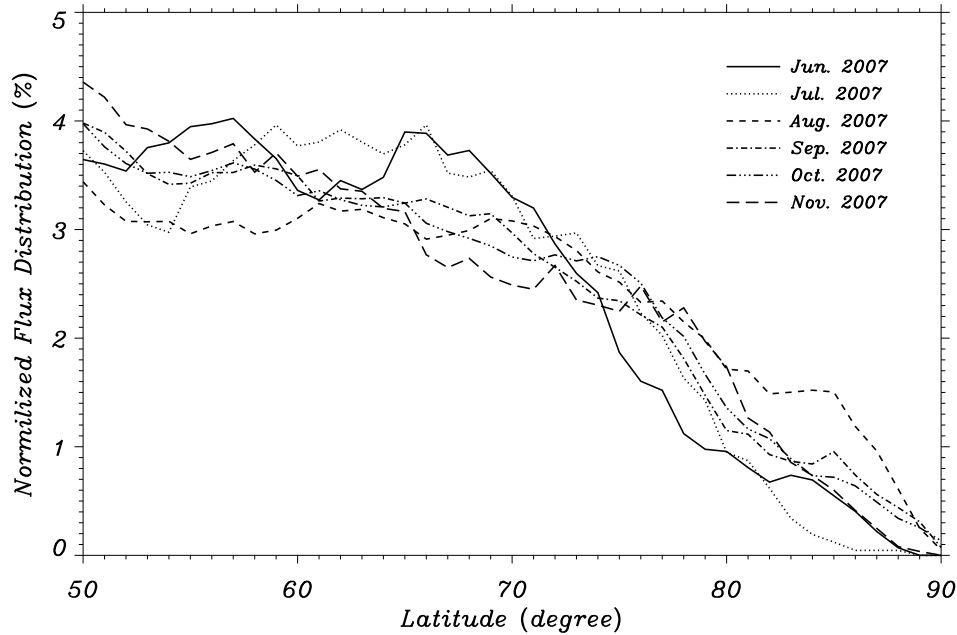}
 \caption{Monthly normalized magnetic flux distribution as a function of latitude for the north
 polar cap from June through November 2007. Distributions from individual magnetograms have been
 normalized by the observable area surface curve. The tilt angle of the solar axis, $B_0$, is
 taken into account. Smoothing by 3 bins of latitude has applied to the different curves to
 reduce the statistical noise in the distribution. 
\label{svsm2007_8542_norm_flux_dist_jun_nov}}
 \end{figure}

The normalized flux element density is plotted in Figure~\ref{svsm2007_8542_norm_flux_dist_jun_nov}
for each monthly period from June through November 2007. The histogram obtained from every
magnetogram is divided by the area surface curve computed taking into account the corresponding
tilt angle, $B_0$, of the polar axis of the Sun. This is to remove any bias in the distribution
that might be due to the decrease of the observable surface area with latitude. The correspondence
between the different curves and time is displayed on the same Figure. The curves are smoothed by 3
bins of latitude in order to reduce statistical variations. It is noticeable how the overall
variation trend is similar for all the different curves

The distribution density of polar flux elements normalized by the surface area is relatively
constant from mid-latitudes up to about 65$^\circ$-70$^\circ$ with a slight increase in the
equatorial direction. Beyond latitude 70$^\circ$ the decrease in the density distribution is
significant with a nearly constant slope to the pole. These results are in complete agreement with
previous ones of the period of time spreading from September to December of 2006 (see Raouafi,
Harvey, \& Henney 2007) showing that flux concentration elements are more abundant near the edge
of the polar cap than near the solar pole. Bearing in mind that flux concentrations form the base
of coronal polar plumes, our results are also compatible with ones found by Raouafi, Harvey \&
Solanki (2007) showing that polar plumes would preferentially be rooted away from the solar pole by
more than 10$^\circ$. 

\section{Conclusion and Discussion}

The high quality of magnetograms from SOLIS allowed us to characterize the flux concentration
distribution as a function of latitude over a relatively large period of time. This was not
possible earlier mainly because of instrumental limitations. The obtained results are important for
studies of magnetic flux transport. They provide additional constraints on solar phenomena such as
meridional circulation that is not possible to measure beyond $45^\circ-50^\circ$ nor how it
functions near solar poles. Meridional flows, for instance, are an important input for solar dynamo
and flux transport models. Our results on the density distribution of the magnetic flux
concentrations at the polar regions suggests that the mechanisms responsible for the flux transport
increasingly lose strength within the last 20 degree latitude before reaching the solar poles.

\acknowledgments
SOLIS data used here are produced cooperatively by NSF/NSO and NASA/LWS. The National Solar
Observatory (NSO) is operated by the Association of Universities for Research in Astronomy, Inc.,
under cooperative agreement with the National Science Foundation. NER's work is supported by NSO
and NASA grant NNH05AA12I.


\begin{thebibliography}{}

\bibitem[Babcock and Babcock(1955)]{Babcock2_55} Babcock, H. W., \& Babcock, H. D. 1955, \apj, 121, 349

\bibitem[Devore \& Sheeley(1987)]{devorsheeley87} Devore, C. R., \& Sheeley, N. R., Jr. 1987,
\solphys, 108, 47

\bibitem[(19)]{} Durrant, C. J., Turner, J. P. R., \& Wilson, P. R. 2004, \solphys, 222, 345

\bibitem[Duvall(1979)]{duvall79} Duvall, T. L. 1979, \solphys, 63, 3

\bibitem[Howard(1974)]{howard74} Howard, R. 1974, \solphys, 38, 59

\bibitem[]{} Jones, H. P., Harvey, J. W., Henney, C. J., Hill, F., \& Keller, C. U. 2002, Proc. IAU
Colloquium 188. Ed. H. Sawaya-Lacoste. ESA SP-505, p.15-18

\bibitem[(19)]{} Keller, C. U., Harvey, J. W., \& Giampapa, M. S. 2003, Proc. SPIE,4853, 194

\bibitem[LaBonte \& Howard(1982)]{labontehoward82} LaBonte, B. J., \& Howard, R. 1982, \solphys, 80, 361

\bibitem[Leighton(1964)]{leighton64} Leighton, R. B. 1964, \apj, 140, 1547

\bibitem[Raouafi, Harvey \& Solanki(2006a)]{rhs06a} Raouafi, N.-E., Harvey, J. W., \& Solanki, S.
K. 2006a, Proc. SOHO-17 Workshop. Eds. H. Lacoste \& L. Ouwehand.
ESA SP-617. Published on CDROM, p.16.1

\bibitem[Raouafi, Harvey \& Solanki(2006b)]{rhs06b} Raouafi, N.-E., Harvey, J. W., \& Solanki, S.
K. 2006b, Proc. IAUS 233. Eds. V. Bothmer \& A. Abdel Hady. Cambridge: Cambridge University Press,
2006., pp.193-194

\bibitem[Raouafi, Harvey \& Solanki(2007)]{rhs07} Raouafi, N.-E., Harvey, J. W., \& Solanki, S. K.
2007, \apj, 658, 643

\bibitem[Raouafi, Harvey \& Henney(2007)]{rhs07} Raouafi, N.-E., Harvey, J. W., \& Henney, C. J.
2007, \apj, 669, 636

\bibitem[Saito (1958)]{saito58} Saito, K. 1958, \pasj, 10, 49

\bibitem[Sheeley et al. (1987)]{sheeleyetal87} Sheeley, N. R., Jr., Nash, A. G., \& Wang, Y.-M.
1987, \apj, 319, 481

\end{thebibliography}
\end{document}